\documentclass[a4paper,onecolumn,11pt]{quantumarticle}
\pdfoutput=1
\usepackage[utf8]{inputenc}
\usepackage[english]{babel}
\usepackage[T1]{fontenc}
\usepackage{amsmath}
\usepackage{hyperref}
\usepackage{braket}
\usepackage{tikz}
\usepackage{lipsum}
\usepackage[numbers]{natbib}

\usepackage{graphicx}% Include figure files
\usepackage{dcolumn}% Align table columns on decimal point
\usepackage{bm}% bold math
\usepackage{color} % change color of text

\newcommand{\beq}{\begin{equation}}
\newcommand{\eeq}{\end{equation}}
\newcommand{\beqa}{\begin{eqnarray}}
\newcommand{\eeqa}{\end{eqnarray}}

\begin{document}
\title{Invariant-based inverse engineering for balanced displacement of a cart-pole system}
\author{I. Lizuain}
\affiliation{Department of Applied Mathematics, University of the Basque Country UPV/EHU, Bilbao 48013, Spain}
\affiliation{EHU Quantum Center, University of the Basque Country UPV/EHU, 48940 Leioa, Spain}
\orcid{0000-0001-9207-4493}
\author{A. Tobalina}
\affiliation{Department of Applied Mathematics, University of the Basque Country UPV/EHU, Bilbao 48013, Spain}
\affiliation{EHU Quantum Center, University of the Basque Country UPV/EHU, 48940 Leioa, Spain}
\orcid{0000-0002-6100-0720}
\author{A. Rodriguez-Prieto}
\affiliation{Department of Applied Mathematics, University of the Basque Country UPV/EHU, Bilbao 48013, Spain}
\affiliation{EHU Quantum Center, University of the Basque Country UPV/EHU, 48940 Leioa, Spain}
\orcid{0000-0002-9030-0060}

\maketitle
\begin{abstract}
Adiabaticity is a key concept in physics, but its applications in mechanical and control engineering remain underexplored.
Adiabatic invariants ensure robust dynamics under slow changes, but they impose impractical time limitations.
Shortcuts to Adiabaticity (STA) overcome these limitations by enabling fast operations with minimal final excitations.
In this work, we set a STA strategy based on dynamical invariants and inverse engineering to design the trajectory of a cart-pole, a system characterized by its instability and repulsive potential.
The trajectories found guarantee a balanced transport of the cart-pole within the small oscillations regime. 
The results are compared to numerical simulations with the exact non-linear model to set the working domain of the designed protocol.
\end{abstract}

\section{Introduction}
Underactuated systems, characterized by having fewer control inputs than degrees of freedom, are prevalent in a multitude of applications, ranging from robotics and aerospace to biomechanics \cite{Spong1998}. These systems present significant control challenges due to inherent nonlinearities and limited actuation capabilities. Consequently, the development of effective control strategies for these systems, which are crucial for many economically relevant areas, has been an active field of research over the last decades.

The cart-pole system has become a widely recognized benchmark for evaluating control strategies for underactuated systems. This inverted pendulum setup, consisting of a pole hinged to a cart moving along a horizontal track (see Fig. \ref{scheme_fig}), has been intensively studied in control theory because of its direct applications in various areas such as personal transport devices \cite{Nikpour2020}, robotic manipulators used in assamply lines \cite{Siciliano2016} and humanoid robots, where the cart-pole model helps develop balancing algorithms \cite{Huang2001}. Moreover, it is a usual benchmark for reinforcement learning algorithms \cite {Wang2020} and neouromorphic computing \cite{Plank2025}. Recently, the cart-pole model has even been extended to the quantum realm, where classical machine learning techniques have been used to balance the system.  Here we take the same path but in the opposite direction.  We apply control techniques originally developed for quantum systems to the classical cart-pole system.

The concept of adiabaticity is ubiquitous in physics, but it is not fully exploited in control engineering. Adiabatic theorems set the existence of approximate adiabatic invariants, such as the action integral in classical mechanics,  when the control parameters vary slowly enough in time  \cite{Sakurai1993}. ``Shortcuts To Adiabaticity'' (STA) is a set of control methods developed to reach the same results of an adiabatic protocol in short times  \cite{review2013,Odelin2019Review}.  Adiabaticity is often used to drive systems  in a robust manner. 
An example is a load hanging from an overhead crane. If the motion is slow enough, the energy of the pendulum is an adiabatic invariant and stays constant. In particular, if the load starts at the minimum energy configuration, this state is preserved throughout the operation \cite{Torrontegui2017,Gonzalez-Resines2017,Lizuain2020Double}. STA accelerate the process and still achieve the same result, avoiding issues related to long operation times such as accumulation of random and/or uncontrollable perturbations.

STA methods have been succesfully applied to a variety of control operations with quantum systems: 
quantum computation \cite{sarandy2011,palmero2017,delcampo2012,takahashi2017},
cooling \cite{onofrio2017, Dann2020},
quantum transport \cite{torrontegui2011,bowler2012},
quantum state preparation \cite{chen2010b,bason2012,zhang2013,zhou2017}, 
manipulation of cold atoms \cite{torrontegui2012,rohringer2015,schaff2010,schaff2011,torrontegui2013,kiely2013} 
or control of polyatomic molecules \cite{mashuda2015}.
Perhaps surprisingly, because of the differing orders of magnitude involved, the dynamics of the cart-pole is closely related to that of a microscopic particle transport in moving traps \cite{torrontegui2011,Tobalina2017}. In both domains the linearized models imply the displacement of an inverted harmonic oscillator.

Cart pole control tasks usually correspond to either swing-up operations, where the pole goes from a downwards vertical posicion to the unstable equilibruim point at the vertical upwards position, or to balance operations, where the goal is to mantain the system at the upwards position. Here, besides aiming for balance, we consider an specific dynamical task, the displacement of the cart such that the cart-pole starts in the upwards vertical positions at certain location, and it finishes at a different position in the same balanced configuration. The design of the control operation is far from trivial since we will be dealing with a repulsive potential with unstable equilibrium \cite{Lanlan_Entropy2023}. Our STA approach is presented here without feedback but it may be combined with feedback control techniques.

The article is organized as follows. The physical model and Hamiltonian of the system are set in Sec. \ref{sec_model}, both in exact form and in the small oscillation regime. In Sec. \ref{STA_sec} dynamical invariants are identified and the STA protocol is designed.
Numerical results are presented in Sec. \ref{results_sec} and, finally, in Sec. \ref{sec_conclusions} we end with the  conclusions and discuss some open questions.

\section{Physical model and basic equations}
\label{sec_model}

\begin{figure}[h!]
\centering
\includegraphics[width=5cm]{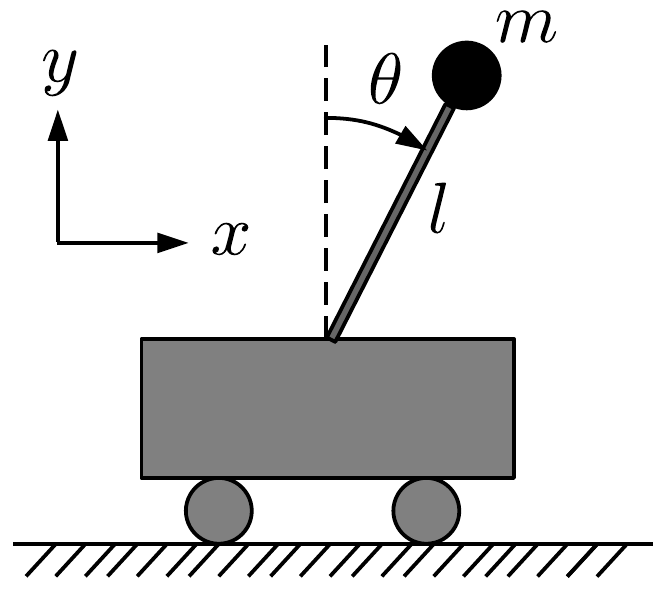}
\caption{Inverted pendulum on a cartpole. Physical model and relevant parameters.}
\label{scheme_fig}
\end{figure}

The physical model and relevant parameters are shown in Fig.  \ref{scheme_fig}.
The model assumes several conditions and idealizations: 
{ (i)} the mass of the wires and friction are neglected;
{ (ii)} point masses; 
{ (iii)} constant wire lengths $l$; 
{ (iv)} the cartpole position is treated as a control parameter rather than a dynamical
variable. 
This last assumption is a common simplification \cite{Abdel-Rahman2003}, often used to facilitate control design, but it requires a 
well-designed controller to ensure accurate implementation. 
A more fundamental approach, where the cartpole position is treated as a dynamical variable, is also be possible and
would allow for a more detailed analysis of energy consumption of STA protocols as explored in \cite{Torrontegui2017}.

In terms of the generalized angle $\theta$, the position of the load is given by its cartesian coordinates
\begin{eqnarray}
 x(t)=x_0(t)+l\sin\theta(t)\textrm{ ; }y(t)=l\cos\theta(t)
\end{eqnarray}
where the cartpole's position $x_0(t)$ is the control parameter to be engineered. The kinetic ($T$) and potential ($V$) are written as
\begin{eqnarray}
 T&=&\frac{1}{2}m\left(\dot x^2+\dot y^2\right);\quad V=mgy,
\end{eqnarray}
where dots represent time derivatives and where we have dropped the explicit time dependence from the variables for simplicity.
The Lagrangian of the system can then be written as $\mathcal{L}=T-V$, from where the equation of motion of the load 
can be easily derived from Euler-Lagrange equations 
$\frac{d}{dt}\frac{\partial \mathcal{L}}{\partial \dot\theta}-\frac{\partial \mathcal{L}}{\partial\theta}=0$ ,
\begin{eqnarray}
\label{eom_exact}
l\ddot\theta -g\sin\theta =-\ddot x_0\cos\theta.
\end{eqnarray}
Using now the horizontal displacement of the mass $q=\sin\theta$ as the new variable and assuming small oscillations, the dynamics of the system is described, to first order in $\theta$, by the linear equation
\begin{eqnarray}
\label{eom_approx}
\ddot q-\omega^2 q=-\ddot x_0
\end{eqnarray}
which corresponds to a forced inverted harmonic oscillator with natural frequency $\omega^2=g/l$. It is easy to show that 
this equation of motion may also be derived from the Hamiltonian
\begin{equation}
\label{hamiltonianq}
 H=\frac{p^2}{2m}-\frac{1}{2}m\omega^2q^2+m\ddot x_0 q,
\end{equation}
where in this small oscillation regime $p=m\dot q$. 
This approximation linearizes the dynamical equations of motion of the system. Results found with exact and approximate dynamics will be compared later to check the validity of the approximation and its limits.

%%%%%%%%%%%%%%%%%%%%%%%%%%%%%%%%%%%%%%%%%%%%%%%%%

\section{Shortcuts to adiabaticity}
\label{STA_sec}
\subsection{Dynamical invariant}
A dynamical invariant of a Hamiltonian system remains constant during the time evolution  \cite{Damour2000}.
If $I$ is an invariant of $H$, the following equation is satisfied,
\begin{eqnarray}
 \frac{dI}{dt}&=&\partial_t I+\{I,H\}=0,
\end{eqnarray}
where $\{I,H\}$ refers to the Poisson bracket.
Quadratic Hamiltonians with a linear in position term, form the so-called Lewis-Leach  family of Hamiltonians 
whose quadratic invariants are explicitly known \cite{LewisLeach1982}. The invariant $I$ of the Hamiltonian of our 
particular system needs some minor modifications from the Lewis-Leach invariants \cite{LewisLeach1982,Odelin2019Review} due
to the repulsive behaviour of the potential involved. 
In particular, let us take the following expression for $I$,
\begin{eqnarray}
\label{invar}
I&=&\frac{1}{2m}\left(p-m\dot\alpha\right)^2-\frac{1}{2}m \omega^2 \left(q-\alpha\right)^2,
\end{eqnarray}
whose time derivative
\begin{eqnarray}
 \frac{dI}{dt}&=&\partial_t I
 +\frac{\partial H}{\partial p}\frac{\partial I}{\partial q}-\frac{\partial H}{\partial q}\frac{\partial I}{\partial p}
 =\left(m \dot \alpha-p\right) \left(\ddot\alpha-\omega ^2 \alpha+\ddot x_0\right)
 \end{eqnarray}
 will be identically zero as long as the function $\alpha(t)$ satisfies the auxiliary equation
 \begin{eqnarray}
\label{newton_eq}
 \ddot\alpha-\omega^2\alpha=-\ddot x_0.
\end{eqnarray}
Therefore, if $\alpha(t)$ satisfies above equation, $I$, as defined in Eq. (\ref{invar}), will be a dynamical invariant of the Hamiltonian (\ref{hamiltonianq}) and will remain constant during the time evolution of the system.
This auxiliary equation (\ref{newton_eq}) is the Newton equation of motion for a forced harmonic (and inverted) oscillator. 
This $\alpha$ function may be regarded as an auxiliary function satisfying the same Newton equation (\ref{eom_approx}). 
However, we shall impose to $\alpha$ some extra boundary conditions that will guarantee zero final excitations 
as we shall see.

\subsection{Shortcut to adiabaticity}

Let us now impose the following boundary conditions (BC) for the auxiliar equation just derived:
\begin{eqnarray}
\label{BCs_alpha}
\alpha(t_b)=\dot \alpha(t_b)=\ddot \alpha(t_b)=0.
\end{eqnarray}
where $t_b=0,t_f$ stands for \emph{boundary} times. These boundary conditions guarantee that the invariant 
$I$ in Eq. (\ref{invar}) coincides with the Hamiltonian $H$ at initial and final times. 
Moreover, at boundary times, the imposed BCs also imply that $\ddot x(t_b)=0$ so that the Hamiltonian 
represents the total mechanical energy of the system. Therefore, if the above BCs are satisfied, the invariant $I$, Hamiltonian $H$
and total mechanical energy $E$ will coincide at boundary times, $I(t_b)=H(t_b)=E(t_b)$.
Then, if a fast finite-time process is designed so that the auxiliary function $\alpha$ satisfy boundary conditions in (\ref{BCs_alpha}), 
the energy at initial and final times will coincide regardless of the initial conditions since
\begin{eqnarray}
E(0)=H(0)=I(0)=I(t_f)=H_q(t_f)=E(t_f).
\end{eqnarray}

In the following, we will show how to construct the cartpole trajectory $x_0(t)$, by means of an inverse engineering approach, so that the desired conditions in (\ref{BCs_alpha}) are satisfied.

%%%%%%%%%%%%%%%%%%%%%%%%%%%%%%%%%%%%%%%%%%%%%%%%%%
\subsection{Inverse engineering}

The inverse engineering strategy is relatively simple. From (\ref{newton_eq}) and integrating twice with respect to time we may write
the cartpole trajectory as
\begin{equation}
\label{x0_from_nw}
 x_0(t)=\int_0^t \int_0^{\tau_2}\left[\omega^2\alpha(\tau_1)-\ddot\alpha(\tau_1)\right]d\tau_1d\tau_2.
\end{equation}
This function should also obey some physical boundary conditions. In particular, we must impose at final time that
\begin{eqnarray}
\label{BCs_x0f}
 x_0(t_f)=d\textrm{ ; }\dot x_0(t_f)=0
\end{eqnarray}
i. e., the transported distance and the final time smooth operation. There are, in summary, eight boundary conditions to be satisfied:
the six conditions for $\alpha(t)$ given in (\ref{BCs_alpha}), plus these final times conditions (\ref{BCs_x0f}).
Note that the conditions at initial times $x_0(0)=\dot x_0(0)=0$ do not have to be imposed, since they are automatically satisfied by construction.

We adopt a seventh-degree polynomial ansatz for the auxiliary function $\alpha$:
\begin{eqnarray}
 \alpha(t)=\sum_{k=0}^7a_k \tau^k
\end{eqnarray}
where $\tau=t/t_f$. The eight free parameters $a_k$ will be determined by imposing the corresponding boundary conditions.
This ansatz is one possible choice among many functional forms. However, we select this polynomial ansatz for its simplicity, favorable mathematical properties (such as smoothness and ease of differentiation) and its flexibility in efficiently satisfying boundary conditions.

\begin{figure}[t]
\centering
\includegraphics[width=15cm]{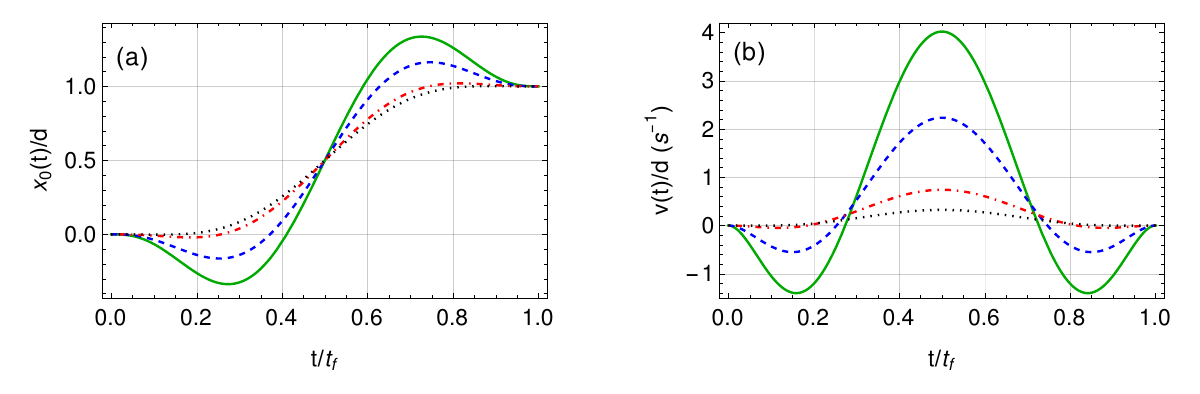}
\caption{(Color online) Cartpole trajectories (a) and velocities (b) for different total process time $t_f$. 
$t_f=1.5$s green-solid, $t_f=2$s dashed-blue, $t_f=4$s red dashed-dotted and $t_f=8$s black dotted. 
Other parameters: $l=1$m.}
\label{trajectories_fig}
\end{figure}

Paramateres $a_0,\hdots,a_5$ are obtained imposing (\ref{BCs_alpha}), while the remaining $a_6, a_7$ are obtained imposing (\ref{BCs_x0f}). This leads to the following cartpole trajectory,
\begin{eqnarray}
\label{x0_tr_long}
 x_0(t)&=&
 d\left[
 126 \tau ^5
 -420 \tau ^6
 +540 \tau ^7
 -315 \tau ^8
 +70 \tau ^9\nonumber\right.\\
 &+&\left.\frac{1}{\omega^2t_f^2}\left(-2520 \tau ^3+12600 \tau ^4-22680 \tau ^5+17640 \tau ^6-5040 \tau ^7\right)\right]
\end{eqnarray}
see some cartpole trajectories and velocities in Fig. \ref{trajectories_fig}.
For short times, the velocities and accelerations involved in the process are quite large as expected, and there are several segments of braking and acceleration. On the other hand, for slow processes ($\omega t_f\gg 1$), the 
second term in (\ref{x0_tr_long}) vanishes and the trajectory $x_0(t)$ tends asymptoticcally to
\begin{equation}
 x_\infty(t)=d\left(126  \tau^5-420  \tau^6+540  \tau^7 -315  \tau^8 +70  \tau^9\right).
\end{equation}
In this regime, there is only one acceleration time segment up to $t_f /2$ (inflexion point of $x_\infty$) and a subsequent braking segment. The maximal velocity reached by the cartpole in ths asymptotic scenario is 
$\dot x_\infty(t_f/2)=\frac{315}{128}\left(\frac{d}{t_f}\right)$.

%%%%%%%%%%%%%%%%%%%%%%%%%%%%%%%%%%%%%%%%%%%%%%%%

\section{Numerical Results}
\label{results_sec}

\begin{figure}[h]
\centering
\includegraphics[width=16cm]{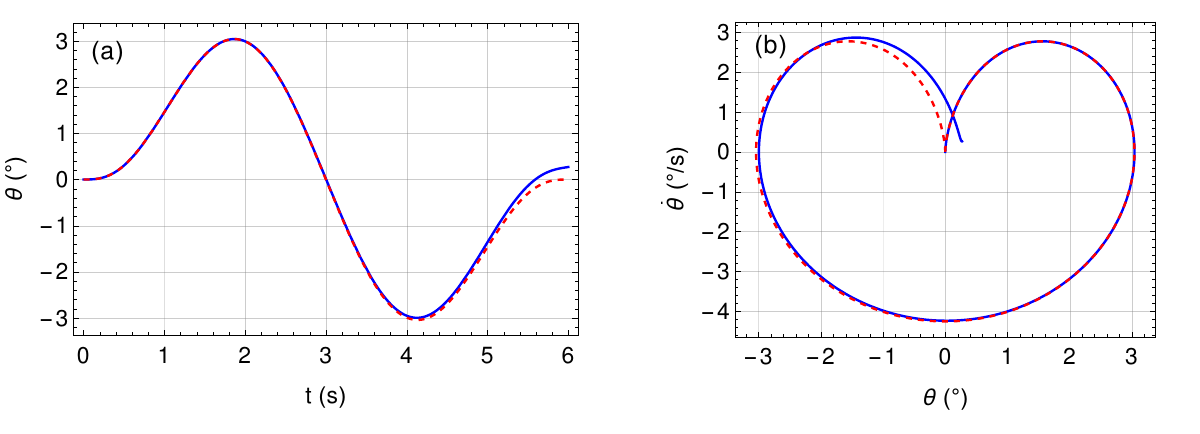}
\caption{(Color online) 
(a) Cartpole swing angle $\theta$ as a function of time after numerically integrating the exact 
dynamical equation (\ref{eom_exact}) (blue solid line) and linearized harmonic model (\ref{eom_approx}) (red dashed line).
with null initial conditions $\theta(0)=\dot\theta(0)=0$. (b) Phase-space diagram for the same process. 
Linearized model leads to purely periodic motion and therefore closed orbits in phase-space, whereas nonlinearities of the exact, more realistic and complex model, lead to deviations in the ideal target configuration at final time.
Other parameters: $l=10$m, $d=2$m, $t_f=6$s.}
\label{theta_evolution_fig}
\end{figure}

Once the trajectory for the cartpole has been designed, the dynamical equation of motion can be integrated either numerically in its 
\emph{exact} form (\ref{eom_exact}) or analytically in its approximated, harmonic linearized version (\ref{eom_approx}). 
See some results in Fig. \ref{theta_evolution_fig}.

It is important to underline that the designed shortcut protocol will work as long as the small oscillation regime holds, i. e., 
as long as the swing angle $\theta$ is small throughout the transport process.
We shall only consider the case where the cartpole is initially in equilibrium (null initial conditions), since this is the most
interesting scenario. In this situiation, a \emph{perfect} shortcut should lead 
to the same final configuration (pendulum in equilibrium). This is indeed what happens if the linear equation of motion is integrated 
(red-dashed lines in Fig. \ref{theta_evolution_fig}). In the exact case however, deviations from the target final configuration are observed (blue solid lines in Fig. \ref{theta_evolution_fig}).
For large transport distances $d$, short pendulum lengths $l$ or short process times $t_f$,
anharmonic effects become more and more important, deviating from the ideal result and limiting the validity of
our shortcut protocol. We study these non-linear effects in detail in the following sub-section.

\subsection{Non-linear effects}

Let us start by looking at the contribution that the different physical parameters have on the the non-linear effects in the dynamical equation (\ref{eom_exact}) when the cartpole trajectory given by (\ref{x0_tr_long}). The role of the transport distance $d$ and pendulum length $l$ is clear. Larger distances $d$ imply larger accelerations so an increase of anharmonic effects is expected and,
a larger pendulum, implies a \emph{flatter} potential so the non-linear effects would decrease as $l$ increases.

The effect of the transport total time $t_f$ is not that clear at first sight. This quantify enters the differential equation as a parameter via the cartpole's acceleration term on the right hand side of (\ref{eom_exact}). It is then clear that longer processes
would imply lower accelerations, so that non-linear effects would decrease as $t_f$ increase. But, on the other hand, $t_f$ also enters the dynamical equation as the final integration limit. Therefore, long processes would also imply an increase of the anharmonic effects since for longer integration times these effects would start to accumulate, leading to undesired final tiempo excitations.

This is more clearly visualized if we write (\ref{eom_exact}) using a dimensionless time $\tau=t/t_f$ as a parameter. 
The equation  of motion (that should be now integrated from $\tau=0$ to $\tau=1$) takes the dimensionless form
\begin{eqnarray}
\label{dimensionless_eom}
 \ddot \theta  &=& \frac{g t_f^2}{l} \sin\theta - \frac{d}{l} a_1 \cos\theta - \frac{d}{g t_f^2}a_2 \cos\theta,
\end{eqnarray}
where dots now represent derivatives with respect to $\tau$ and 
$a_{1,2}$ are different terms of a \emph{dimensionless} acceleration
\begin{eqnarray}
a_1(\tau)&=&5040 \tau ^7-17640 \tau ^6+22680 \tau ^5-12600 \tau ^4+2520 \tau ^3\\
a_2(\tau)&=&-211680 \tau ^5+529200 \tau ^4-453600 \tau ^3+151200 \tau ^2-15120 \tau.
\end{eqnarray}
The effect of each parameter is now clear: larger $d$ or smaller $l$ implies larger non-linear effects, whereas the effect of $t_f$ is twofold, so a compromise in this magnitude would be needed.

\begin{figure}[t!]
\centering
\includegraphics[width=15cm]{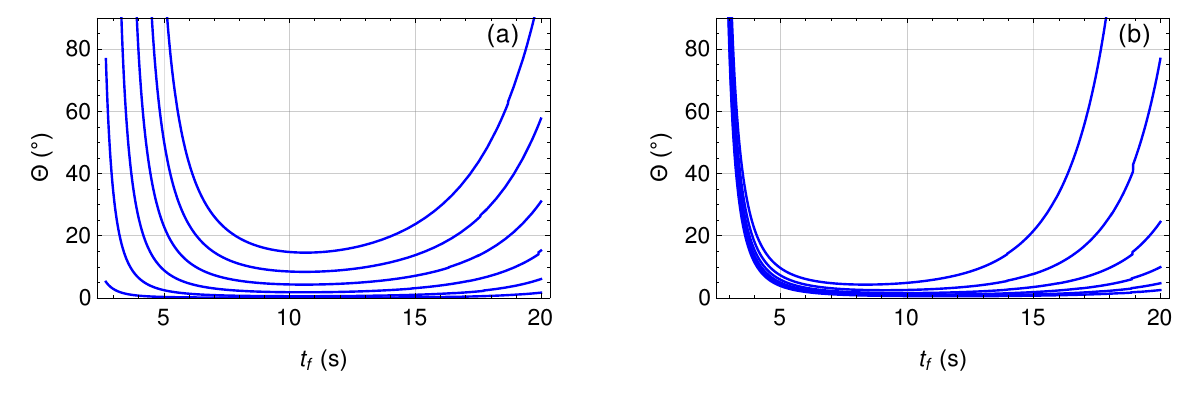}
\caption{(Color online) Ficticious angle $\Theta$ as a function of the process time $t_f$ in different scenarios. 
As discussed in the text, this angle quantifies the final deviation from the ideal balanced final state.
Numerical results clearly indicate that both excessively fast or slow processes result in undesirably high final angle configurations, whereas an optimal time window minimizes final excitation.
 Outside this optimal region, nonlinear effects become increasingly significant, causing the protocol to break down and leading to final configurations with very large angles.
(a) From bottom to top $d=2, 4, 6, 8, 10, 12$m for a fixed pendulum length of $l=16$m.
(b) From bottom to top $l=10, 12, 14, 16, 18, 20$m for fixed value of $d=5m$.
Pendulum initially in equilibrium.}
\label{fict_angle_fig}
\end{figure}

\begin{figure}[t!]
\centering
\includegraphics[width=8cm]{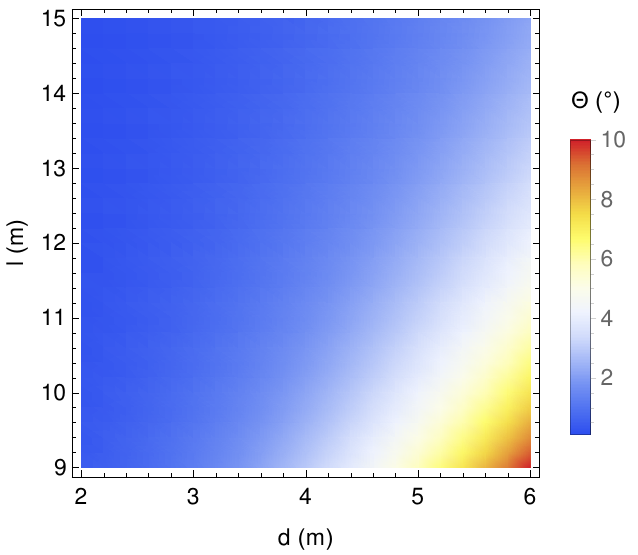}
\caption{(Color online) Minimum of final time excitations measured by the value of the ficticious angle $\Theta$ (in º) as a function of $d$ and $l$. 
For each $d$-$l$ configuration, the optimal process time $t_f$ is numerically calculated as the minimum of each curve in Fig. \ref{fict_angle_fig}. Using this optimal time, the possible minimum fictitious angle is computed for that particular configuration. In other words, this plot illustrates the best-case scenario for a given $d$-$l$ configuration.
For larger values of $d$ or smaller values of $l$, final excitations larger than $10^\circ$ are obtained.}
\label{Theta_min_fig}
\end{figure}

To quantify the excitation at final time in a way that is easy to understand and visualize, we
measure the final energy $\Delta E$  in terms of a \emph{fictitious} angle $\Theta$. This angle is defined
as the final angle when the final energy is considered (artificially) to be purely potential.
This angle is a way te measure the final excitation energy in terms of a more visual quantity.
Taking the pendulum vertical position as the zero of the potential energy  and considering null initial conditions (pendulum initially in
equlibrium), the initial energy is $E(0)=0$, so the energy difference is given by
\begin{eqnarray}
\Delta E=E(t_f)-E(0)=\frac{1}{2}m l^2\dot\theta(t_f)^2 + 2 m g l \sin^2\frac{\theta (t_f)}{2},
\end{eqnarray}
where the values of $\theta$ and $\dot\theta$ at $t=t_f$ are obtained by numerically integrating (\ref{eom_exact}).
This final energy has contributions from both kinetic and potential energies. One can artificially consider this final 
energy to be purely potential by defining the ficticious angle $\Theta$ by the relation 
$\Delta E=2 m g l\sin^2\frac{\Theta}{2}$ so that
\begin{eqnarray}
 \Theta=2 \arcsin \left[\sqrt{\frac{l \dot\theta(t_f)^2}{4 g}+\sin ^2\frac{\theta(t_f)}{2}}\right].
\end{eqnarray}
Some results of this quantity are shown in Fig. \ref{fict_angle_fig} in different scenarios. 
As previously commented, for a given configuration of transport distance $d$ and pendulum length $l$, too short or too long processes
lead to undesired final excitations.
One can also numerically obtain the optimal value of each process time $t_f$ 
so that the resultant ficticious angle $\Theta$ is minimal for a given $d$-$l$ configuration (minima of each curve in Fig. \ref{fict_angle_fig}), see some results and comments in Fig. \ref{Theta_min_fig}.

%

%%%%%%%%%%%%%%%%%%%%%%%%%%%%%%%%%%%%%%%%%%%%%%%%
\section{Conclusions}
\label{sec_conclusions}

In this work, an invariant based inverse engineering STA method has been applied  to control the transport of a mass on an inverted pendulum cartpole, a mechanical system characterized by its instability and repulsive potential. The designed transport protocol 
minimizes final energy excitations without requiring feedback. 
Numerical simulations using the exact non-linear model allows us to check the parameter interval where the protocol is accurate and works properly, and also allows us to quantify the final excitations due to its inherent non-linear behaviour. These simulations validate the effectiveness of our approach, demonstrating its potential for practical applications in robotics and automation.

The results highlight the advantages of STA methods in overcoming the limitations of adiabatic processes, which, while robust, impose impractically long operation times. The inverse engineering approach allows us to generate trajectories that achieve fast and minimal
excitation transport, expanding the applicability of STA beyond quantum and microscopic systems to classical mechanical engineering problems.

One of the key features of the STA approach we have followed is its flexibility. It should be underlined that we have not really optimized the trajectory. We have inverse engineered the trajectory by means of a polinomial ansatz, but of course this choice, and therefore the solutions to the inverse problem are not unique, leaving room for further optimization based on specific performance criteria, such as robustness to different kind of perturbations and/or energy cost and efficiency. 

Although this work focuses on non-feedback control, hybrid control strategies combining STA with some feedback mechanism is an interesting open line for future research.
Future work could also explore the extension of this protocol to more complex systems such as dynamical systems with higher degrees of freedom (double or triple pendulums).
These extensions would further improve the robustness of the method to real-world industrial and robotic applications.

\section*{Acknowledgments}
This research was funded by the Basque Government through Grant No. IT1470-22 and  
Spanish Government MCIU through Grant No. PID2021-126273NB-I00.

%%%%%%%%%%%%%%%%%%%%%%%%%%%%%%%%%%%%%%%%%%%%%%%%%%%%%%%%%%%%%%%%%%
% \bibliographystyle{quantum}
% \bibliographystyle{unsrt}
% \bibliography{mybib}

% \bibliographystyle{unsrt}
% \input{cartpole_arxiv.bbl}

\end{document}